# Thermal Dendrites on the Surface of Water and Water Solution

D. A. Rusova[1, a)] and L. M. Martyushev[1, 2, b)]

[1]*Ural Federal University, Mira St. 19, 620002 Ekaterinburg, Russia.*
[2]*Institute of Industrial Ecology, S. Kovalevskoi St. 20a, 620219 Ekaterinburg, Russia.*

a) dariarusova@mail.ru
b)leonidmartyushev@gmail.com

**Abstract.** Thermal dendrites (fractal-like structures) on the surface of water and some water solutions are found with an infrared camera. They are observed with specific sizes and temperature differences in the liquid and are not associated with the movement of the liquid.

## 1. INTRODUCTION

A convection associated with the inhomogeneous heating of liquids and gases is one of the most common processes in nature. Rayleigh-Benard convection: convection in a thin horizontal layer of liquid heated from below turns out to be very convenient for experimental studies [1]. Modern matrix infrared cameras make it possible to simply study local temperature gradients on the surface of the liquid with great accuracy [2-4]. The structures observed on the surface of a liquid is influenced by many factors: the physicochemical properties of liquids, temperature gradients, hydrodynamic flows, cell geometry, impurities, etc. [1-4]. This provides a huge field for research. The complexity of the mathematical analysis of such hydrodynamic structures (mostly nonlinear [1]) necessitates experimental exploratory research in various directions in order to better understand the phenomenon.

The purpose of this work was to study using an infrared camera thermal structures on the surface of water (and a number of aqueous solutions) in small cells with an open surface at temperatures close to room temperature.

## 2. EXPERIMENTAL METHOD

The experiment used a round glass container (Petri dish) with a diameter of 102 mm with an open top surface. It placed liquid. The height of the liquid (h) ranged from 1 to 20 mm. The container was placed on a horizontal massive temperature-controlled heater. Its temperature varied from 30 to 45°C, the ambient temperature was from 24 to 27°C. The studied liquids: pure and distilled water (electrical resistance, respectively 6 MOm·cm and 1 MOm·cm), drinking water, 0.9% aqueous solution of NaCl (saline), saturated aqueous solution of NaCl, aqueous solutions of sucrose or glycerin of various concentrations.

For the measurement, the infrared camera (Infratec ImageIR 8800) was used. It have a spectral range of (8 ... 9.4) μm, a temperature resolution of 0.035 ° C and a matrix of 640*512. To control the temperature, in addition to the infrared camera, the temperature of the liquids was measured using thermocouples.

# 3. RESULTS AND DISCUSSION

We list below the most interesting observed results, which are new in comparison with previous studies (see Refs. [2-4]).

1. On the surface of the water (regardless of the degree of its purity), as well as on the surface of the investigated aqueous solutions, an approximately similar pattern is observed with the infrared camera. At first, for some time a lot of thin threads (with an approximate thickness of 1-2 mm) are seen, they form irregularly shaped figures (bundles) and change in time (Fig. 1).

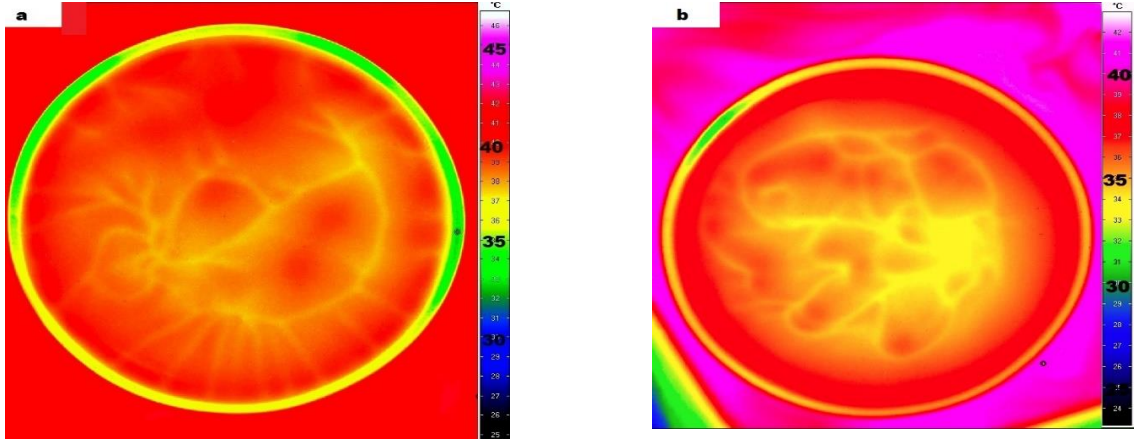

**FIGURE 1.** The bundle of threads (the distribution of temperature on the surface of distilled water according to the infrared camera). (a) h=10 mm, (b) h=5 mm. The temperature scale is shown on the right

2. Then dendritic (fractal-like) structures are very often formed from them (Fig. 2). These structures (let's call them DF) have a pronounced center from which numerous branches "grow", the number of which increases to approach the edges of the experimental container.
3. These DF structures for the temperature range studied (container temperature 30-45°C, ambient temperature 24-27°C) are observed when the thickness of the liquid in the container is about from 2 to 10 mm. The Rayleigh number is in the range $10^3$-$10^4$, and the Marangoni number is about $10^4$. With increasing liquid thickness, the structure on the surface becomes similar to that shown in Fig. 3. With a decrease in the liquid thickness, the structure on the surface looks like Fig.4. These two last structures (Fig.3.4) were previously observed in Refs.[2-4]. Note that DF structures (for a liquid thickness from 2 to 10 mm) were also not observed in a container with a much smaller size (diameter is 25 mm).
4. The temperature of the branches of DF structures is smaller than the temperature of the liquid between the branches in about 2-3°C. This difference approximately corresponds to the horizontal temperature difference between the center and the edge of the liquid in the container. When air is blown over the surface of the liquid, the temperature difference (between the branches and their surrounding) increases further. This indicates the importance of evaporation from the surface for the formation of DF structures. The central upper part of the container is always the coldest area, and the edges of the container (due to the proximity of the side walls) are the hottest area.
5. If the liquid does not heat up and has an average ambient air temperature, then a small temperature gradient arises on the surface due to evaporation. As result, DF structures can appear. In this case the temperature of the branches of DF structures is smaller than the temperature of the liquid between the branches in about 0.3–0.5°C. However, these structures are less pronounced and less stable.
6. DF structures have exclusively thermal origin and are located in a thin layer on the surface. The addition of small particles (tracers) shows that there are no visible movements on the surface of the liquid. This is also noted in Refs.[3,4]. A mechanical movement (with the help of thin metal wire) of the liquid layers below the thin surface layer does not affect the DF structure. These structures, once formed, are fairly stable. Indeed, small mechanical perturbations of DF structure only temporarily "destroy" them. If the surface of the liquid is rotated, the DF structures rotate as well.

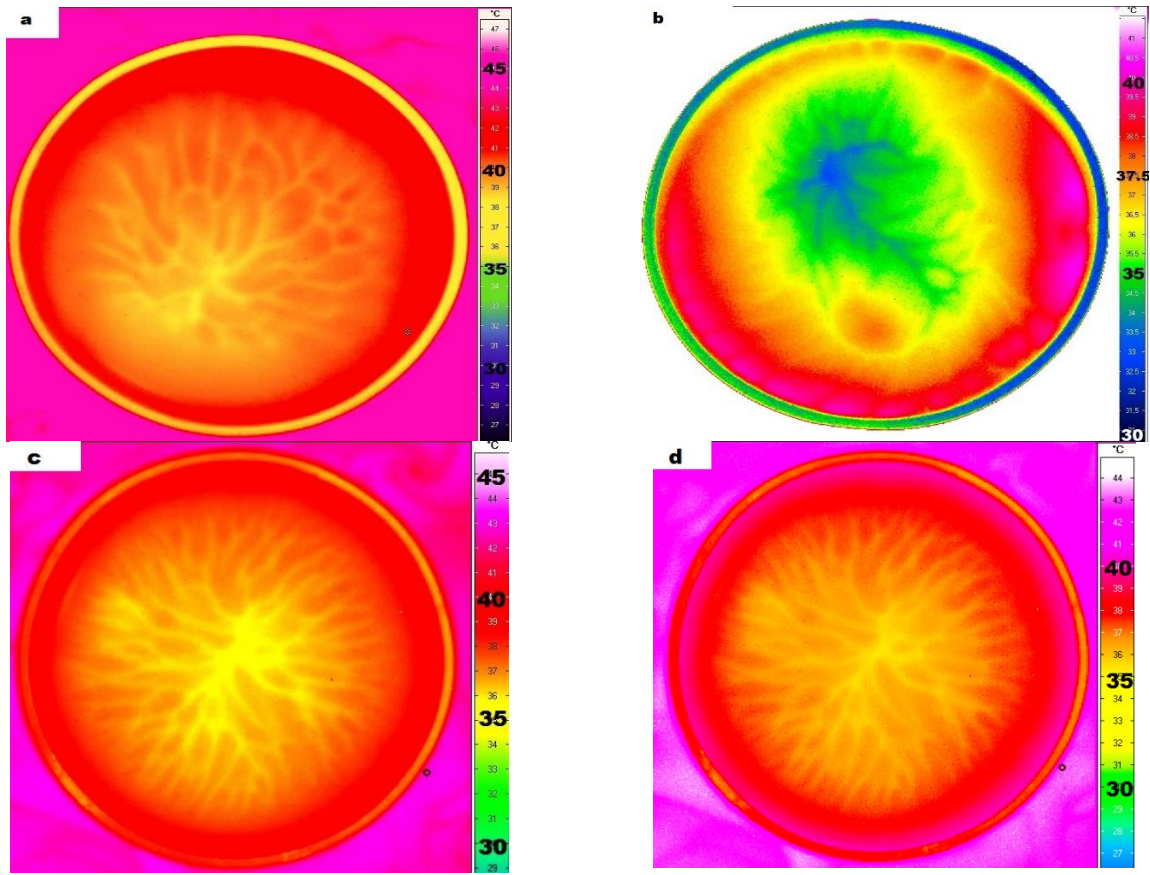

**FIGURE 2.** Asymmetric (a,b) and symmetric (c.d) dendrites (fractal-like structures, DF structures). This is the distribution of temperature on the surface of liquid according to the infrared camera. (a) saline, h=5 mm, (b) saline, h=10 mm, (c) saturated aqueous solution of NaCl, h=10 mm, (d) saturated aqueous solution of NaCl, h=5 mm. The temperature scale is shown on the right

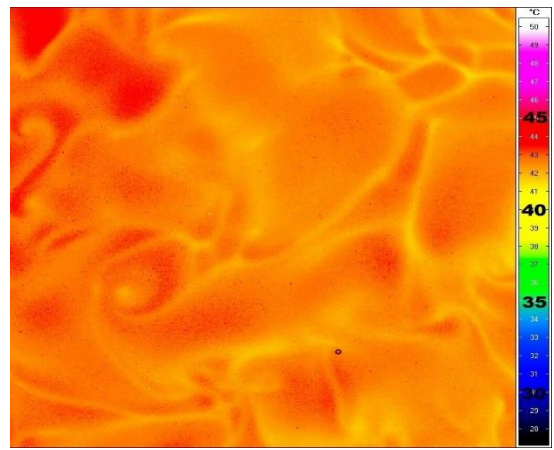

**FIGURE 3.** Mosaic-vortex structure (the distribution of temperature on the surface of drinking water according to the infrared camera). h=200 mm. The temperature scale is shown on the right

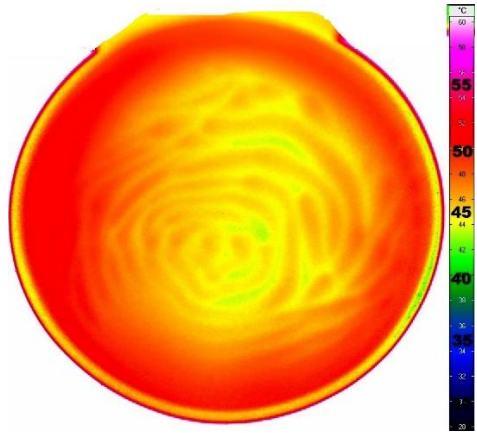

**FIGURE 4.** Rolls (the distribution of temperature on the surface of distilled water according to the infrared camera). h=1 mm. The temperature scale is shown on the right

7. Under the above temperature conditions and thickness (sizes) of the liquid in the container, pure glycerin does not produce thermal dendritic structures on the surface (Fig. 5).

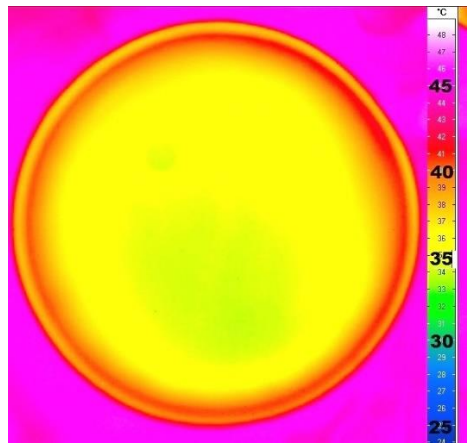

**FIGURE 5.** The distribution of temperature on the surface of pure glycerin according to the infrared camera. h=5 mm. The temperature scale is shown on the right

So, the main result is the detection of thermal (non-hydrodynamic) DF structures on the water surface. These structures arise in the presence of a temperature gradient on the surface in a thin surface layer of the liquid. The specific thin layer of water is apparently the cause of the appearance of DF structures. Indeed, in the absence of a similar layer (such as in solids or glycerin), the temperature equalization on the surface is possible through heat conduction without the formation of special structures. If this layer is disturbed (even partial) due to thermogravitational convection (for a large thickness of the liquid in a container) or due to thermocapillary convection (for a small thickness of the liquid), horizontal temperature difference is equalized (not local, but integral, because there is a (quasi-)periodic alternation of warm and cold areas on the surface) and the DF structure do not arise.

In this special thin layer of water, structures arise with a very large interface between the "cold" and "warm" regions. Due to such self-organization, the nonequilibrium system most effectively equalizes the temperature between the central region and the edges of the container. Nature "selects" a fractal branched structure as the most efficient way for heat transfer. The surface "radiator" self-organizes. The physical nature of this "radiator" must be investigated in the future with the help of additional experiments and numerical modeling. Although, apparently, hydrodynamic flows are absent in this layer, the thermophysical system remains rather complicated. Indeed, it is a thin layer of water (tens of micrometers) with numerous differently distributed impurities, the boundary conditions on different sides of this layer are significantly different and must take into account all the mechanisms of heat transfer, including evaporation and fluid motion, respectively, on the upper and lower its surfaces. In addition, the influence of thermoelectric phenomena on the appearance of DF structures cannot be excluded [5].

## 4. CONCLUSION

Thermal surface tree structures found in the work were not noted earlier in the literature known to the authors. This paper presents their preliminary study. The nature of thermal dendrites and the domain of their existence is of great interest and requires further study.